\documentclass{article}

\usepackage{PRIMEarxiv}

\usepackage[utf8]{inputenc} % allow utf-8 input
\usepackage[T1]{fontenc}    % use 8-bit T1 fonts
\usepackage{url}            % simple URL typesetting
\usepackage{booktabs}       % professional-quality tables
\usepackage{amsfonts}       % blackboard math symbols
\usepackage{nicefrac}       % compact symbols for 1/2, etc.
\usepackage{microtype}      % microtypography
\usepackage{lipsum}
\usepackage{fancyhdr}       % header
\usepackage{graphicx}       % graphics
\graphicspath{{media/}}     % organize your images and other figures under media/ folder
\usepackage{mdframed}
\usepackage{xcolor}
\usepackage{subcaption}
\usepackage{float}
\usepackage{listings}
\usepackage{soul}
\usepackage{hyperref}
\usepackage{xr}
\usepackage{amsmath}
\usepackage{makecell}
\usepackage{diagbox}
\makeatletter

\usepackage[backend=biber, style=nature, citestyle=numeric-comp, sorting=none, maxbibnames=6]{biblatex}
\renewbibmacro*{in:}{%
  \iffieldundef{journaltitle}
    {}
    {\printtext{\bibstring{in}\intitlepunct}}}

\usepackage{subcaption}

\title{Unlocking Multi-Task Electric Energy System Intelligence: Data Scaling Laws and Performance with Limited Fine-Tuning}

\author{
  Shaohuai Liu\thanks{Equal contribution as joint first co-authors}, ~Lin Dong\footnotemark[1], 
  \textbf{Chao Tian}\\
  Department of Electrical and Computer Engineering\\
  Texas A\&M University \\
  College Station, Texas, USA\\
  \And
  Le Xie (Corresponding author)\\
  John A. Paulson School of Engineering and Applied Sciences\\
  Harvard University\\
  Cambridge, Massachusetts, USA\\
  \url{xie@seas.harvard.edu} \\
}

\begin{document}
\maketitle

\begin{abstract}\label{abs:main}
Data scaling has revolutionized research fields like natural language processing, computer vision, and robotics control, providing foundation models with remarkable multi-task and generalization capabilities. In this paper, we investigate whether similar data scaling laws exist in developing foundation models for power systems, and whether appropriate data scaling can yield multi-task, cross-timescales capabilities that can be deployed in \textit{unseen} operational scenarios. To this end, we conducted a comprehensive empirical study on data scaling by fine-tuning open-source foundation models using labeled data collected from diverse operational tasks and scenarios. We study how a foundation model's scenario generalization performance evolves with the number of training tasks, scenarios, and demonstrations. Our study involved collecting more than 450k demonstrations and implementing independent tests under a rigorous evaluation framework. Our findings reveal several key insights: First, the generalization performance of a fine-tuned foundation model follows an approximate power-law relationship with the number of demonstrations and scenarios.  
Second, the fine-tuned model also demonstrates impressive multi-task capabilities, where multi-task training shares similar performance improvements with single-task training as the number of demonstrations increases, without interference among tasks.
Lastly, models with small parameter sizes could have strong performance as well. Model performance does not scale significantly with parameter size. These findings underscore the feasibility of developing multi-task foundation models tailored for power systems, demonstrating that while larger datasets and models generally improve performance, extreme scaling is unnecessary to achieve satisfactory outcomes. 
\end{abstract}


\section{Introduction}

Scaling has been a vital motivation for recent artificial intelligence (AI) research surges. In natural language processing (NLP)~\cite{brown2020language}, computer vision (CV)~\cite{radford2021learning, liu2023visual}, and recent advancements in robotics control~\cite{brohan2023rt, kim2024openvla}, many studies have identified scaling laws demonstrating that model performance and generalization capabilities improve with increases in \textit{dataset size}, \textit{parameter number}, \textit{training computation} and \textit{inference computation}~\cite{achiam2023gpt,jaech2024openai,lin2024data}. However, comprehensive scaling laws have not been established in developing foundation models for power systems, which prevents research and industrial practice from following a similar path. In this paper, we explore the first dimension of scaling, data, as scaling data is a prerequisite for scaling parameters and inference computation. We aim to investigate whether data scaling law exists in power systems, especially in the context of multiple operational tasks across different timescales, and if so, what insights and guidance they might provide for building foundation models for power systems.

While data scaling has significantly enhanced models in NLP and CV with exceptional generalization capabilities, most of today's power system AI models still lack comparable generalization and multi-task capability. From this perspective, we prioritize a multi-task foundation model generalizable to \textit{unseen} operational scenarios as the first-class goal. In this context, we seek to address the following fundamental question: \textit{Can appropriate data scaling facilitate the development of power system foundation models that are capable of performing multiple tasks and adapting to diverse scenarios?}

To explore this, we conduct a comprehensive empirical study on data scaling by fine-tuning a foundation model for power systems, which is a predominant methodology for adapting foundation models to domain-specific tasks. We define this generalization ability as scenario generalization, where the scenario represents the combination of distinct load, renewable, power generation and system state profiles. Scenario generalization refers to the ability of a trained model to maintain task performance in previously unseen scenarios of the same task. Furthermore, we particularly emphasize such scenario generalization ability in multi-task settings, where the goal is to generalize across different scenarios while handling diverse tasks within the same physical power network. 
% We categorize generalization into two dimensions: \textit{scenario generalization} and \textit{task generalization}, which essentially covers all factors a foundation model may encounter during real-world deployment. 
At this stage, we do not pursue system-level generalization across power networks with different topologies, as we believe that it would require collecting vast amounts of data from thousands of systems, which is beyond the scope of this work. Instead, we systematically explore how a foundation model's multi-task performance evolves in new operational scenarios as the number of scenarios and demonstrations increases within a single power system.

Different from previous research employing foundation models to either \textit{generate text-only response}~\cite{majumder2024exploring} or \textit{generate codes to operate tools}~\cite{huang2024large, jia2024enabling}, we leverage the foundation model to directly generate responses to task queries, which involve floating-number states in both prompts and answers. However, the common tokenizers are typically not designed to accommodate float numbers. Inspired by the success of vision-language-action models~\cite{brohan2023rt,kim2024openvla} to generate float actions in responses, we adopt similar designs on float number discretization and tokenization, allowing the foundation model to adapt to both textual and float number inputs effectively. In addition, rather than focusing solely on specific tasks such as power flow analysis or predictions~\cite{hamann2024foundation,tu2025powerpm, liu2024lfllm}, we emphasize the multi-task ability within a single model, which could be critical to various real-world power system applications.

We examined our findings under controlled simulation, which is commonly accepted in power system research. Our experiments are based on the IEEE-118 standard test case, which offers greater complexity and realism compared to previous toy attempts~\cite{majumder2024exploring,huang2024large}. We employ broadly used simulation and optimization tools to collect training demonstrations, which are structured as question-answer (QA) pairs (see Sec.\ref{sec:multi-task-dataset} for details). Operational scenarios are generated using historical generation and load profiles. To evaluate scenario generalization capabilities, we set up two separate datasets, \textit{training} and \textit{test}. Throughout our research, we collected over 450,000 demonstrations and scenarios and conducted all experiments under a rigorous evaluation protocol that included other independent test scenarios. Our extensive investigation reveals several surprising results:

\begin{itemize}
    \item \textbf{Simple power laws}. The generalization ability to \textit{unseen} scenarios scales approximately as a power law with the amount of training data.
    \item \textbf{Multi-task scalability}. The data scaling power law can extend to multi-task models, with negligible interference among tasks and no significant performance degradation. 
    % \item \textbf{Performance surge using limited data}. Even with a limited amount of training data, foundational models can achieve satisfactory performance. This capability opens up diverse training strategies for users with varying accuracy requirements. 
    \item \textbf{Efficient data requirement}. Satisfactory performance can be achieved with a relatively small amount of training data compared to other domains, suggesting that power system foundation models can be efficiently developed with limited-sized datasets. 
        
    \item \textbf{Parameter-scaling with limited impact}.  In task-limited application scenarios, parameter scaling does not always lead to performance improvements due to potential parameter redundancies. This insight highlights a promising direction for the efficient training of demand-oriented foundation models.
    % \item \textbf{Foundation behavior is easier than expected}. Collecting data in as many tasks as possible, each with one unique physical system and 50 thousand demonstrations, allows training a model with foundation behavior that generalizes well to new scenarios across all trained tasks. 
\end{itemize}

\section{Investigation of Data Scaling Laws}
In this section, we first explore how increasing the number of demonstrations affects the scenario generalization capability of foundation models under single-task training. Next, we investigate if the scenario generalization capability could hold under multi-task training. Throughout all experiments, we found a power-law in data scaling property in fine-tuning power system foundation models. 
% Based on these, we further demonstrate efficient data collection and training strategies to develop a multi-task capable foundation model for power systems. 

\subsection{Results and Quantitative Analysis}
\begin{figure}
    \centering
    \includegraphics[width=1.0\linewidth]{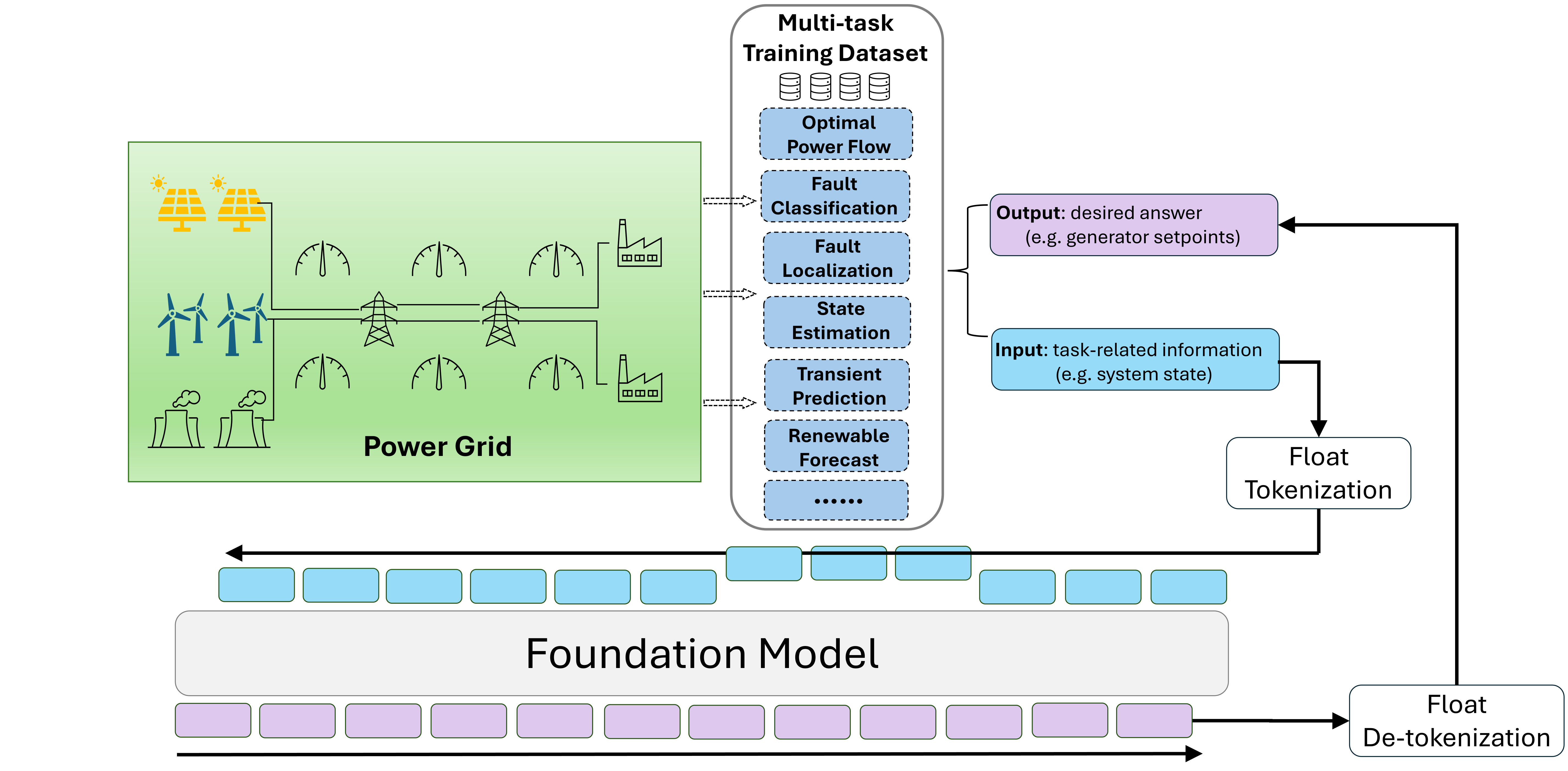}
    \caption{\textbf{Overview of the research.} The training data used in the research were either collected from real world historical data or generated through well-recognized powers system simulation tools, such as Powerworld, Pypower, etc. A hybrid dataset covering typical power system operation tasks was then formed and converted into Question \& Answer pairs. After specific transformation of float numbers in the question data, the model were fine-tuned to generate expected answers, which were finally converted back to appropriate form through the same transformation algorithm.}
    \label{fig:overview}
\end{figure}

\textbf{Tasks}.
We first focus on three typical operational tasks in power systems: \textit{Optimal Power Flow}, \textit{Fault Classification} and \textit{Fault Localization}. In \textit{Optimal Power Flow}, the foundation model is supposed to take state information of buses, generators, and branches as inputs, then generate the best active power setpoints of the generators. In \textit{Fault Classification}, the model is supposed to identify the fault type based on the dynamics of system states prior to and after a fault. In \textit{Fault Localization}, the model is supposed to predict the exact location where the fault occurs given the system dynamics. Illustrations of the answer generation process for all tasks are shown in Fig.\ref{fig:overview}.

%, with further details available in Appendix.
%\textcolor{red}{TODO: Add an overview figure like OpenVLA}

\textbf{Scenario generalization}. For \textit{Optimal Power Flow}, we use up to 90,000 load, generation and renewable profiles with 5-minute actual intervals to generate corresponded QA demonstrations. Power flow simulations are conducted to determine system states based on the given load, generation, and renewable profiles. The system states and profiles at the next timestep are then formulated as question prompts, and we utilize the Pypower OPF toolbox to generate the ground-truth answers to the OPF problems. For \textit{Fault Classification} and \textit{Fault Localization}, similarly, we use up to 100,000 load and generation profiles to get steady power flow solutions, and randomly assign one of five distinct faults to a randomly selected bus or branch to simulate diverse fault scenarios\cite{zheng2022multi}. Hence there are 678 unique fault scenarios in total.

\begin{figure}
    \centering
    \includegraphics[width=1.0\linewidth]{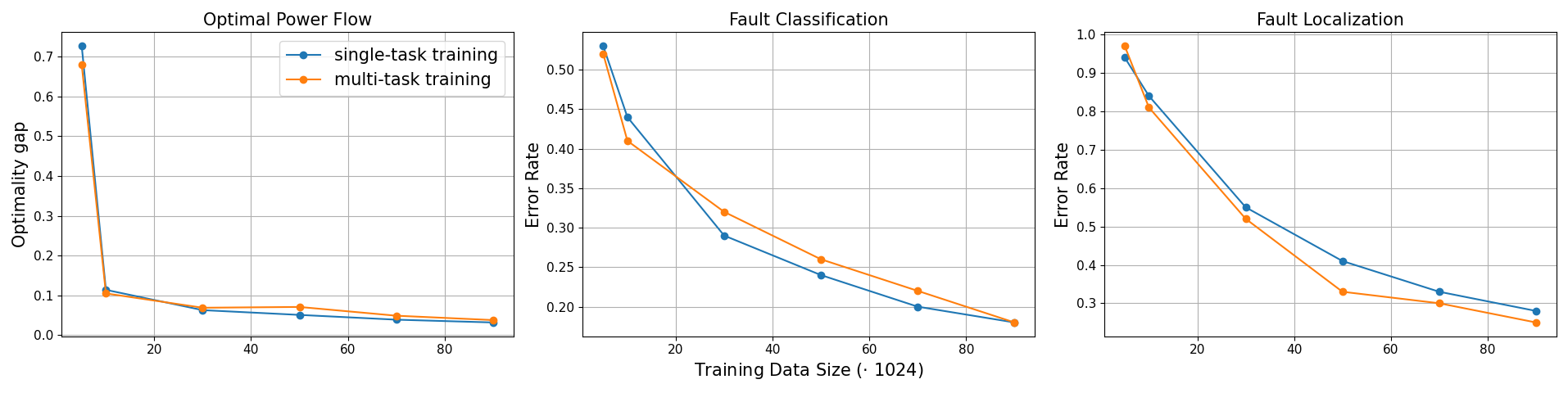}
    \caption{\textbf{Single-task training vs. Three-task training}. Orange lines represent curves under single-task training, while blue lines are under three-task training. The closer the two lines are, especially in the convergence stage, the stronger the multi-task ability, demonstrating the task scalability of foundation models in the power system.}
    \label{fig:single-multi-1-3}
\end{figure}

The orange curves in Fig.\ref{fig:single-multi-1-3} present the results for the three proposed tasks under single-task training, leading to a key observation: as the amount of training data increases, the model's performance on unseen scenarios (with previously unseen load and renewable profiles) almost consistently improves across all fractions of demonstration data. 

\textbf{Multi-task capability}. To investigate whether the scenario generalization capability could extend to multi-task settings, we construct a multi-task hybrid dataset by combining the same single-task datasets used previously to perform multi-task training. This results in a hybrid dataset with 270k QA pairs in total. We adopt the same dataset splits as in single-task training to evaluate the task performance changes under multi-task training settings. In this sense, we maintain the same amount of training data for each task as in single-task setting.

The blue curves in Fig.\ref{fig:single-multi-1-3} illustrate the results for the three proposed tasks when fine-tuned simultaneously on the hybrid dataset. The results reveal a notable pattern: multi-task simultaneous training does not degrade the model performances compared to single-task training, which indicates that the foundation models could be task-scalable. To further validate this argument, additional tasks will be introduced and analyzed in following sections.

\subsection{Power-law fitting}
We then explore whether our experimental results follow power-law scaling laws as observed in foundation models for NLP, CV, and Robotics~\cite{kaplan2020scaling,lin2024data}. Mathematically, if two variables $X$ and $Y$ satisfy $Y=\alpha X^k$ then they fit a power-law relationship. 
A linear relationship will be revealed if applying a logarithmic transformation to both $X$ and $Y$
\begin{equation}
    \log(Y)=k\log(X)+\log(\alpha)
\end{equation}

\begin{figure}
    \centering
    \includegraphics[width=1.0\linewidth]{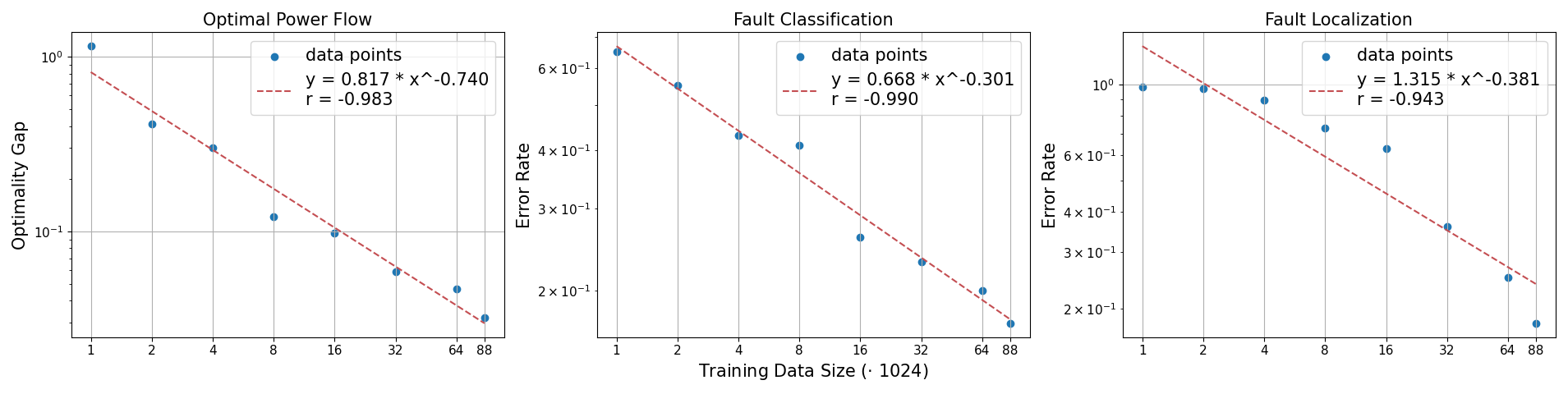}
    \caption{\textbf{Power-law relationship}. Dashed lines represent power-law regressions, with fitted equations provided in the legend. All axes are presented in logarithmic scales. The correlation coefficient r indicates how strong a power-law holds between the scenario-generalization ability in each task and the training data size.}
    \label{fig:power-1-3}
\end{figure}

In power system context, Y could represent metrics such as convergence rate, optimality gap, error rate, and mean squared error. X denotes the number of demonstrations or scenarios. To examine this relationship, we fit a linear model to the log-transformed data, as shown in Fig.\ref{fig:power-1-3}. Based on all the results, we summarize the following data scaling laws:
\begin{itemize}
    \item The foundation model's generalization capability to new scenarios scales approximately as a \textit{power-law} relationship with the amount of data used in the fine-tuning.
    \item The data scaling law
    holds not only in single-task but also in multi-task settings, suggesting that fine-tuning foundation models for power systems can be task-scalable.
\end{itemize}

These power-law relationship regarding the number of demonstration-scenario pairs and its invariance to multiple tasks can serve as predictive tools for developing a multi-task capable and cross-timescale foundation model for power systems.

\section{Verification of task scalability}
To verify the task scalability of the data scaling behavior, we add new tasks into the fine-tuning process and assess whether the power-law relationship between scenario generalization performance and demonstration amounts can still be achieved and whether the strong multi-task capability can be preserved as the number of tasks increases. We extend our analysis by incorporating three new tasks: \textit{Transient Prediction}, \textit{Renewable Prediction}, and \textit{State Estimation}. 
In \textit{Transient Prediction}, the model is supposed to predict the system dynamics after a fault occurs. The model is provided with a period of system dynamics prior to the occurrence of the fault, along with fault information, including the fault type, location, and time. 
In \textit{Renewable Prediction}, the model is supposed to predict the minute-level wind and solar maximum generation for the next hour, given the weather forecasts (including temperature, sunlight, wind speed, etc.) for the same period. In this task, we also provide the historical renewable generation curves from the previous hour in the prompt.
In \textit{State Estimation}, power flow simulations are performed on the 90,000 renewable and generation profiles, with scaled normal noise added to the results to simulate real-world system measurements. The model is supposed to estimate the bus voltage magnitudes and phase angles based on these noisy measurements.

% \textcolor{red}{There should be another paragraph that describes the results with 3 added-on tasks}.

\begin{figure}[htbp]
    \centering
    \begin{subfigure}{1.0\textwidth}
        \centering
        \includegraphics[width=\linewidth]{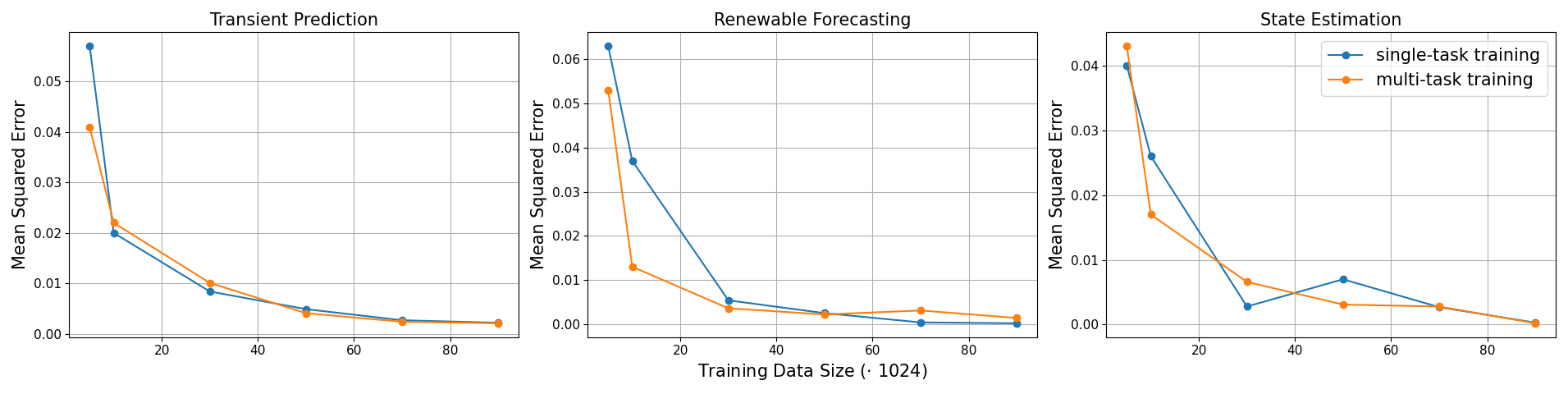}
        \caption{Single-task vs. Multi-task training.}
        \label{fig:single-multi-4-6}
    \end{subfigure}

    % \vspace{0.5cm} 

    \begin{subfigure}{1.0\textwidth}
        \centering
        \includegraphics[width=\linewidth]{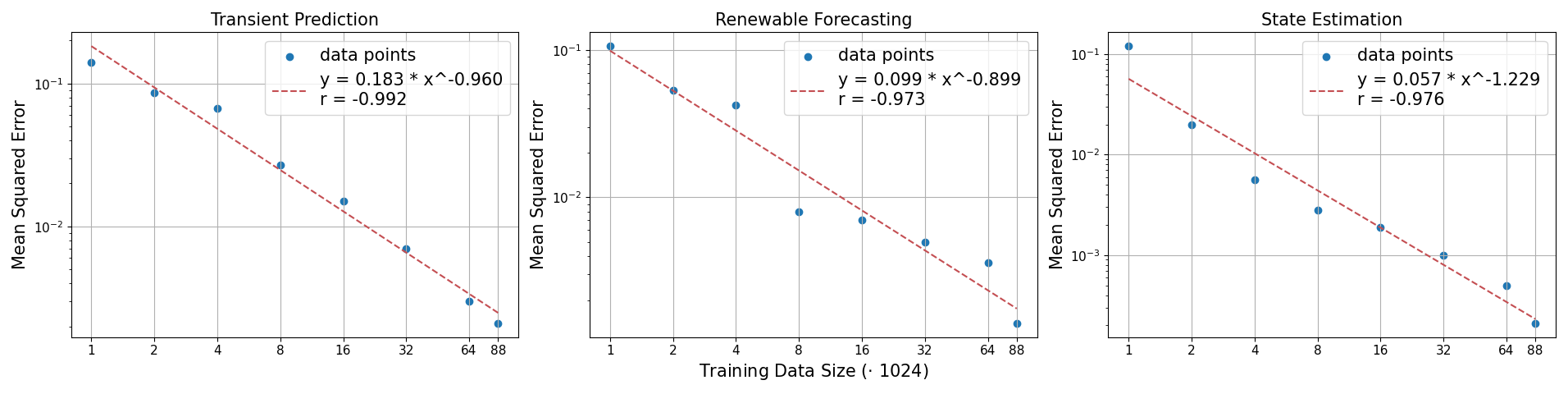}
        \caption{Power-law fitting.}
        \label{fig:power-4-6}
    \end{subfigure}

    \caption{\textbf{Verifications on three new tasks.} To verify the effectiveness and scalability of our proposed method, we extend the number of simultaneous training tasks from 3 to 6. We only demonstrate the results of three new-added tasks since the model performs the same for the three previous tasks. As demonstrated in (A), the model still maintains the same multi-task performance compared to single-task training when the number of simultaneous training tasks increases. In addition, the model performance still follows the power-law relationship with the data size in the three new tasks, as depicted in (B).}
\end{figure}

Consistent with the previous three-task setting, the results as shown in Fig.\ref{fig:single-multi-4-6} and Fig.\ref{fig:power-4-6} further reinforce our previous key findings: with appropriate data collection, the foundation model's performance will be well preserved compared to single-task training. The data scaling law remains valid even as the number of tasks doubles, demonstrating the model’s robust scalability across multiple tasks.

% \section{Scenario diversity: efficient data collection}

% \begin{table}[h]
%     \centering
%     \begin{tabular}{|c|c|c|c|c|c|c|}
%     \hline
%         \diagbox{Task}{Seasons} & \multicolumn{2}{|c|}{1 season} & \multicolumn{2}{|c|}{2 seasons} & \multicolumn{2}{|c|}{4 seasons} \\
%         \hline
%         Data amount & 1000 & 2000 & 1000 & 2000 & 1000 & 2000 \\
%         \hline
%         % OPF convergence rate$\uparrow$ &  &  &  \\
%         % \hline
%         OPF optimality gap$\downarrow$ & 1.150  & 0.414 & 0.330 &  &  & \\
%         \hline
%         Fault detection accuracy$\uparrow$ & 35\% & 45\%  &  &  & & \\
%         \hline
%         Transient prediction MSE$\downarrow$ & 0.1399 & 0.086  &  &  & & \\
%         \hline
%         Renewable prediction MSE$\downarrow$ & 0.084 & 0.054 &   &  & &\\
%         \hline
%         State estimation MSE$\downarrow$ & 0.123 & 0.021 &  & 
%  & & \\
%         \hline
%     \end{tabular}
%     \caption{\textbf{Parameter scaling experiments}. 1B, 3B, and 8B are the numbers of parameters for different model sizes. All models are trained with 64k QA pairs for each task under the multi-task training setting. Scaling parameters has very limited influences on the final converged performance.}
%     \label{tab:scenario-diversity}
% \end{table}

\section{Model size: beyond data scaling} 

Finally, we extend our exploration beyond data scaling to investigate the model parameter scaling. We use the Llama 3 series models and evaluate the impact of parameter scaling by testing models with 1B, 3B, and 8B parameters. These specific model sizes were chosen due to computational constraints, as processing long-context prompts associated with aforementioned tasks already significantly increases resource demands. The results presented in Table.\ref{tab:parameter-scaling} reveal several interesting observations: (1) \textit{Small models can be strong as well.} Despite their size, the small models demonstrate surprisingly satisfactory performance on each task, which scales predictably with the increasing data size.  (2) \textit{Scaling parameters may not be the most critical factor.} In task-specific power system applications, parameter scaling does not necessarily lead to significant performance gains, likely due to potential parameter redundancies. In addition, we record the training time per loop to demonstrate how computational efficiency changes with the number of model parameters. These findings highlight a promising direction for the efficient training of demand-oriented foundation models. Rather than indiscriminately increasing model size, focusing on optimizing the utilization of existing parameters and improving data quality may yield better results with fewer resources.

% Our investigation also investigates the impacts of selections on training strategies like LoRA. We conduct comparisons between using/not using LoRA, and what layers to fine-tune. Results were shown in \textcolor{red}{Table?}, identifying several interesting observations: (1) As proved in other domains by researchers, LoRA can significantly reduce computation cost while maintaining a certain level of performance in power system domains. (2) Selections of layers to be trained is a critical factor to obtain satisfactory outcomes. \textcolor{red}{Table?}As for the six-task setting, \textcolor{red}{(detailed results)...}

\begin{table}[h]
    \centering
    \begin{tabular}{|c|c|c|c|}
    \hline
        \diagbox{Task\&metric}{Model} & Llama-3.2-1B & Llama-3.2-3B & Llama-3.2-8B \\
        \hline
        % OPF convergence rate$\uparrow$ & 90\% & 90\% & \textbf{92\%} \\
        % \hline
        OPF optimality gap$\downarrow$ & 0.047 & 0.036 & \textbf{0.029} \\
        \hline
        Fault detection error rate$\downarrow$ & 25\% & \textbf{24\%} & 25\%\\
        \hline
        Transient prediction MSE$\downarrow$& 0.0068 & 0.0052 & \textbf{0.0042} \\
        \hline
        Renewable prediction MSE$\downarrow$ & 0.0046 & 0.088 & \textbf{0.0038} \\
        \hline
        State estimation MSE$\downarrow$ & \textbf{0.0015} & 0.0021 & 0.0018 \\
        \hline
    \end{tabular}
    \caption{\textbf{Parameter scaling experiments}. 1B, 3B, and 8B are the numbers of parameters for different model sizes. All models are trained with 64k QA pairs for each task under the multi-task training setting. Scaling parameters has very limited influences on the final converged performance.}
    \label{tab:parameter-scaling}
\end{table}

\begin{table}[h]
    \centering
    \begin{tabular}{|c|c|c|c|}
       \hline
       Model & Llama-3.2-1B & Llama-3.2-3B & Llama-3.2-8B \\ \hline
       $T_{loop}$ & 1.12s & 4.37s & 9.96s \\ \hline
       Multipliers & 1x & 3.90x & 8.89x \\ \hline
    \end{tabular}
    \caption{\textbf{Parameter computational efficiency}. 1B, 3B, and 8B are the numbers of parameters for different model sizes. We use 2 A100 GPUs to train the models. The average time per training loop increases as the number of parameters increases, basically follows a linear relationship.}
    \label{tab:parameter_efficiency}
\end{table}

\section{Discussion, Limitations, and Future Works}
The power energy sector is a physics-based, multifaceted field that requires a deep understanding of various disciplines, such as underlying physical principles, optimization theory, and more. How data scaling impacts foundation models' performances in this domain is the primary focus of our study. Unlike traditional AI models, we investigate the effects of data scaling on a simultaneous multi-task basis, which reflects the generalization capabilities of foundation models across diverse tasks in power system applications. Our findings demonstrate that a multi-task, scenario-generalizable foundation model can be achieved with a limited amount of training data and a relatively small base model tailored to a specific power network. This insight also highlights the potential for user-customized foundation models, which can strike a balance between user-specific demands and computational efficiency.

However, our work has several limitations that future research should address. First, our study focuses on data scaling for tasks and scenarios based on a specific power grid topology (in this case, the IEEE 118-bus system) and does not explore topology-level generalization. Achieving such generalization would require data collection from a wide range of diverse power grid topologies or a fundamental restructuring of the base model to adapt to different configurations, which is beyond the scope of this study. Second, due to accessibility constraints, our training data used in this study are primarily synthetic, generated through rigorous simulations but lacking validation with real-world power grid data. Future work could incorporate real grid topologies and operational data to further validate these findings. 
Lastly, we directly employ the question-answer framework in a language format for foundation model processing, which is intuitive, however, brings long-context prompts and insufficient computation efficiency. We encourage researchers to explore architecture innovations that aligns language instructions and power grid data in a more efficient way.
% Lastly, we consider NLP-type tasks in power system domains—such as semantic understanding, text summarization, and Q\&A—to be relatively straightforward to achieve in a general sense. As a result, we did not include such tasks in our experiments. However, in the development of practical foundational models, these tasks could be incorporated into the training dataset to enhance the model's versatility and applicability.

\section{Approach}
In this section, we present the core technologies and design choices adopted in the study, which enable the foundation model to directly analyze and operate power grids through fine-tuning. First, we introduce a specially designed tokenization method that allows LLMs to handle float-rich data problems effeciently. Next, we outline the dataset construction process , which equips the foundation models with multi-task operational capabilities. We then describe the fine-tuning framework and the base model selections. Finally, we introduce the evaluation metrics designed to assess the performance of fine-tuned foundation models on common power system tasks.

\subsection{Float tokenization}
To enable large language models to analyze and control power grids, they must be trained to process float numbers effectively. Inspired by the success of Vision-Language-Action models, we adopt a direct approach to this problem by representing floats as tokens in the model's inputs and outputs, which are treated the same as normal language tokens. Our float encoding is based on the discretization approach proposed by the RT-1 model. Each question-answer pair may include varying sizes of float grid state vectors depending on the tasks. To standardize the representation, we first normalize all floats to the range $[-1,1]$ by dividing them by the maximum absolute value. Then, all the continuous floats are discretized into $B$ bins uniformly, where 0 represents -1 and $B-1$ represents 1. To be more specific, the discretization process is defined mathematically as follows:
\begin{equation}
    v_d=\lfloor(v/|v|_{max}+1)\cdot B/2\rfloor
\end{equation}
where $v_d$ represents the discretized integers and $|v|_{max}$ denotes the maximum absolute value.
We also record these maximum absolute values and use them to convert discretized integers back to continuous floats as follows:
\begin{equation}
    \tilde{v}=(2v_d/B-1)\cdot|v|_{max}
\end{equation}
We illustrate how to discretize the float-rich prompts based on the optimal power flow task, converting a float-rich prompt into a single string by concatenating discretized tokens for each float,
\begin{equation*}
\begin{aligned}
    \text{I will provide you with the input states of buses:} [1.04,\ & 26.1,\ 88.2,\cdots \\
    & \Downarrow\\
    [1025, & 1151, 1314,\cdots
\end{aligned}
\end{equation*}

In order to use these discretized prompts to fine-tune a large language model into a power system foundation model, we need to associate tokens from the model's existing vocabulary with the discretized float bins, which requires reserving $B$ tokens to serve as float tokens. The tokens to choose depend on the specific tokenizer used by each LLM. Unfortunately, for the Llama 3 series we used as the base model, the tokenizer only reserves 100 'special tokens' for tokens newly introduced during fine-tuning, which is insufficient for the $B$ tokens we required for discretization. To address this limitation, we overwrite the $B$ \textit{least used} tokens in the Llama tokenizer's original vocabulary with our $B$ float tokens. Once the floats are processed into a sequence of tokens, the  base language model is trained with a standard next-token prediction objective, evaluating the cross-entropy loss on the predicted float tokens. This enables the fine-tuned LLM to effectively process the continuous system states and variables as a power system foundation model.

\subsection{Multi-task training dataset}
\label{sec:multi-task-dataset}
The goal of constructing the dataset for training a power foundation model is to capture a large diversity of operational tasks and scenarios. Such comprehensive training data enables the final trained model to handle various operational scenarios, even those out of the training dataset, and maintain coherent performance across tasks. Our designed tasks include steady-state tasks like optimal power flow, renewable prediction, and state estimation. In addition, we also train and evaluate the model on transient-state tasks including fault detection and transient prediction. We use the open-sourced PyPower simulator to generate steady-state data. Then, we employ PowerWorld to generate transient-state profiles based on the simulated steady-state snapshots\cite{thayer2020easy}. Each task is represented in the form of question-answer pairs. To better illustrate the question-answer structure, we demonstrate the details in Table.\ref{tab:question-answer}. 

\begin{table}[h]
    \centering
    \begin{tabular}{|c|c|c|}
    \hline
       Task  & Question & Answer \\
        \hline
        OPF & \thead[l]{We are solving an optimal power flow problem. \\The objective is to minimize the operational cost \\while meeting the power balancing and other\\ constraints. I will provide you the input states of buses\\ \{bus\_states\}, states of generators \{gen\_states\}.\\ What are the best power setpoints? }& \thead[l]{The best active power setpoints of \\generators are \{gen\_settings\}}\\
        \hline
        Fault Detection & \thead[l]{The power network has \{n\_bus\} buses:\{bus\_info\}.\\ A fault of \{fault\_type\} occurred at bus \{bus\_fault\}.\\ Given the time series of dynamic voltages across\\ all buses: \{input\}, please tell me the fault type and location.}& \thead[l]{The fault type is \{fault\_type\}, \\occurred at bus \{fault\_bus\} (between \\bus \{fault\_bus1\} and \{fault\_bus2\}).} \\
        
        \hline
        Transient Prediction & \thead[l]{The power network has \{n\_bus\} buses:\{bus\_info\}.\\ A fault of \{fault\_type\} occurred at bus \{bus\_fault\}.\\ Given the first \{input\_len\} time steps of dynamic\\ voltages across all buses: \{input\}, please predict the \\next \{output\_len\} time steps. }& \thead[l]{The next \{output\_len\} time steps across \\\{n\_bus\} buses are: \{output\}} \\
        \hline
        Renewable Prediction & \thead[l]{The following is a renewable generation power prediction \\task. You need to predict future \{future\_hours\} hours \\generation curve according to the weather prediction data. \\ Solar Zenith Angle: \{angle\_data\}, Wind Speed: \{wind\_data\}, \\ Relative Humidity:\{humidity\_data\}, \\Temperature:\{temperature\_data\}. Please predict the following\\ \{X\} points wind power and solar power.} & \thead[l]{The expected wind power is \\\{wind\_predictions\}. The expected \\solar power is \{solar\_predictions\}.}\\
        \hline
        State Estimation & \thead[l]{We are solving a power system state estimation problem.\\ The power network has \{n\_bus\} buses and  \{n\_branch\} \\branches. We will provide you with the measurements\\ of voltage magnitude, active power injections and \\reactive power injections at each bus. Measurements\\ of voltage magnitude: \{voltage\_mea\}. Measurements of \\active power injection: \{active\_mea\}. Measurements \\of reactive power injection: \{reactive\_mea\}.\\ What are the best estimates for the voltage magnitude \\and voltage angle of each bus?} & \thead[l]{Given above measurements, the best \\estimates for voltage magnitude are: \\\{voltage\_mag\}. Best estimates for \\voltage angle are: \{voltage\_angle\}.}\\
        \hline
    \end{tabular}
    \caption{Detailed question-answer designs for each task.}
    \label{tab:question-answer}
\end{table}

\subsection{Fine-tuning method and base-model selection}
Using the proposed multi-task QA dataset, we apply supervised fine-tuning (SFT), by fine-tuning the model with a next-token prediction mechanism and cross-entropy loss to output correct answers. We employ the decoder-only architecture to leverage key-value cache technology, which enhances computational efficiency, especially important for long-context response settings. In addition, to ensure accessibility for the research community, we use the open-sourced Llama3 model as our base model\cite{llama3}.

Since the token lengths of power system operational question-answer pairs are typically over 20,000 in our settings, it is critical to reduce the training computation burden and memory consumption, especially in the context of limited computation resources. To this end, we employ Low-Rank Adaptation (LoRA)~\cite{hu2022lora} to fine-tune our model. While common practice only fine-tunes transformer query and value heads to achieve computational efficiency, we observed this setting could lead to inferior performance in our case. Therefore, we choose to fine-tune the query, key, value, and projection heads to enhance the expression ability of the fine-tuned model.

\subsection{Evaluation metrics}
We conduct rigorous evaluations to ensure the reliability of our findings. First, to evaluate the scenario generalization performance of the fine-tuned model, we assess it separately in operational scenarios with \textit{unseen} load, renewable, and generation profiles. 
Second, we use widely-accepted metrics to evaluate the fine-tuned model in each task as summarized in Table.\ref{tab:metric}. Specifically, the OPF metric designs follow a reinforcement learning environment reward design for power scheduling tasks~\cite{liu2023real}. The OPF convergence rate indicates the percentage of power flow convergence when applying the foundation model's OPF responses to the power grid. The optimality gap is calculated using the score difference between ACOPF optimizer solutions and fine-tuned model responses. The score is defined following a Lagrange function that considers economic objectives and operational constraints. The design details of the score function can refer to Appendix.\ref{sec:opf_metrics}.
\begin{table}[h]
    \centering
    \begin{tabular}{|c|c|}
    \hline
       Task  & Metric design \\
        \hline
        OPF & Weighted average of sub-metrics including bus voltages, line loadings, reactive power, etc.\\
        \hline
        Fault Detection & Error Rate (1 - Accuracy)\\
        \hline
        Transient Prediction & MSE between predictions and ground-truth\\
        \hline
        Renewable Prediction & MSE between predictions and ground-truth\\
        \hline
        State Estimation & MSE between predictions and ground-truth\\
        \hline
    \end{tabular}
    \caption{Evaluation metric designs for each task.}
    \label{tab:metric}
\end{table}

\printbibliography[title={References}]

\appendix
\label{sec:appendix}

\section{Optimal power flow metrics}
\label{sec:opf_metrics}
The metric of optimal power flow response is designed as a weighted average of the following components considering the power system steady-state operational constraints and economic objectives.

\textbf{Line overflow.} 
To ensure the security of power transmission, we establish a reward mechanism that takes into account the line loadings. The current load rate is defined as the ratio of the transmission current to the maximum transmission current, which is limited by the thermal capacity of the line. 
To prevent line congestion and outages, it is desirable for the transmission lines to maintain a reasonable load level. Consequently, we design the reward function for this aspect as follows:
\begin{equation}
    r_{\text{overflow}}=1-\frac{\sum_i\min(\rho_i,1)}{n_{\text{line}}}
\end{equation}
where $\rho_i$ indicates the current load rate of line $i$. $n_{\text{line}}$ represents the line number. 

\textbf{Renewable consumption.} 
To optimize the utilization of renewable energy, a reward function is formulated to encourage the scheduling agent to consume more renewable energy. This reward function is based on the renewable energy consumption rate, which is defined as the ratio of the total power currently consumed from renewable sources to the maximum power generated by renewable sources. A higher renewable energy consumption rate corresponds to a higher reward, thereby incentivizing the agent to maximize the use of renewable energy. This reward is designed as follows:
\begin{equation}
    r_{\text{renewable}}=\frac{\sum_i\text{p}_i}{\sum_i\text{p}_i^{\text{max}}},\quad i\in\text{renewable units}
\end{equation}
where $\text{p}_i$ represents the power output of generator $i$, $\text{p}_i^\text{max}$ indicates maximal power generation capability of renewable generator $i$. The closer $\text{p}_i$ and $\text{p}_i^\text{max}$ are, the higher the renewable consumption reward.

\textbf{Power balancing.} The balanced generator aims to balance the residual power and eliminates discrepancies between power generation and load consumption, while its generation capability is limited. If its power output exceeds its operational boundaries, a power mismatch occurs, which may result in load shedding or blackouts as previously mentioned. To prevent such mismatches and maintain safe operation, we design the reward function as follows:
\begin{gather}
    r_\text{balance} = -\left(\frac{\max(\text{p}_\text{bal}-\overline{\text{p}_\text{bal}},0)}{\overline{\text{p}_\text{bal}}-\underline{\text{p}_\text{bal}}}+\frac{\max(\underline{\text{p}_\text{bal}}-\text{p}_\text{bal},0)}{\overline{\text{p}_\text{bal}}-\underline{\text{p}_\text{bal}}}\right)
\end{gather}
where $\overline{p_\text{bal}}$ and $\underline{p_\text{bal}}$ indicate the upper bound and the lower bound of the balanced generator's active power per step. If the balanced power $p_\text{bal}$ is out of bounds, there would be a penalty.

\textbf{Operating cost.} We formulate a reward function for the operational costs of thermal units while considering the negligible costs of renewable energy generation. Specifically, the operating costs of thermal units are represented as quadratic functions of output power, and additional costs are incurred for the startup/shutdowns of thermal units. As for renewable sources, their operating costs are considered to be negligible as they do not rely on fossil fuels for power production. 
The operating cost reward is designed as follows:
\begin{equation}
    r_\text{cost} = -\frac{\sum_i c_{i,2}\text{p}_i^2+c_{i,1}\text{p}_i+c_{i,0}+\mathcal{I}(\text{s}_i, \text{s}_i^-)c_{\text{on-off,i}}}{Z}
\end{equation}
where $c_{i,2},c_{i,1}$ and $c_{i,0}$ are the second order, first order and constant coefficients of the operation cost of generator $i$, respectively. The coefficients of renewable units are much lower than that of thermal units. $\text{p}_i$ represents the power output of generator $i$. $\text{s}_i$ represents the on-off status of generator $i$, and the $\text{s}_i^-$ is the status 1-step advance. $c_\text{on-off,i}$ is the startup and shutdown costs of generator $i$. $\mathcal{I}(\text{s}_i, \text{s}_i^-)$ is an indicative function that turns to be 1 if $\text{s}_i\neq\text{s}_i^-$, otherwise 0. $Z$ is the normalization factor set as $10^5$ in experiments.

\textbf{Reactive power.} Reactive power plays a vital role in supporting the voltage stability of the power grid. However, the reactive power output capacity of the generators is constrained. While exceeding this limit is not catastrophic, excessive reactive power compensation can significantly increase operational cost. In light of these considerations, we design 
the reactive power reward as follows:
\begin{equation}
    r_\text{reactive}=\exp\left(-\sum_i\left[\frac{\max(\text{q}_i-\overline{\text{q}_i},0)}{\overline{\text{q}_i}-\underline{\text{q}_i}}+\frac{\max(\underline{\text{q}_i}-\text{q}_i,0)}{\overline{\text{q}_i}-\underline{\text{q}_i}}\right]\right)-1
\end{equation}
where $\text{q}_i$ is the reactive power of generator $i$, and $\overline{\text{q}_i}, \underline{\text{q}_i}$ are the upper bound and the lower bound of generator $i$. There would be a penalty if any generator violates its reactive power constraint. 

\textbf{Bus voltage.} In power system operation, it is common practice to limit node voltage magnitudes within the range of 0.95-1.05 per unit. If the voltage magnitude at a node is too low, it can result in a significant increase in the transmission loss of the grid. Conversely, if the node voltage magnitude is too high, it requires more reactive power compensation and may cause the generator's reactive power to exceed its upper limit.
To regulate the node voltage magnitudes within specified ranges, we design the bus voltage reward similarly to the reactive power reward.
\begin{equation}
    r_{\text{voltage}}=\exp\left(-\sum_i\left[\frac{\max(\text{v}_i-\overline{\text{v}_i},0)}{\overline{\text{v}_i}-\underline{\text{v}_i}}+\frac{\max(\underline{\text{v}_i}-\text{v}_i,0)}{\overline{\text{v}_i}-\underline{\text{v}_i}}\right]\right)-1
\end{equation}
where $\text{v}_i$ is the voltage magnitude of bus $i$, and $\overline{\text{v}_i}, \underline{\text{v}_i}$ are the upper bound and the lower bound of voltage magnitude of bus $i$. There would be a penalty if any bus violates its voltage magnitude constraint.

\end{document}